\documentclass[12pt]{article}

\usepackage{times}
\usepackage{a4wide}
\usepackage[final]{graphicx}
\usepackage{amsmath,amssymb}
\usepackage{authblk} 
\usepackage[numbers,compress]{natbib}
\usepackage[margin=1em,font={small},labelfont={sc}]{caption}

  \renewenvironment{thebibliography}[1]{%
    \begin{oldthebibliography}{#1}%
      \setlength{\parskip}{0ex}%
      \setlength{\itemsep}{0ex}%
      \small
  }%
  {%
    \end{oldthebibliography}%
  }

\begin{document}

\title{\LARGE \bf Neutrino Masses and TeV-scale Particles\\
Testable at the LHC\footnote{Presented by I.P. at the Nuclear and Subnuclear Physics - Symposium held at the Croatian Academy of Sciences and Arts, Zagreb, December 13, 2012.} }

\author{Kre\v{s}imir Kumeri\v{c}ki}
\author{Ivica Picek}
\author{Branimir~Radov\v{c}i\'c}
\affil{Department of Physics, University of Zagreb,
              Bijeni\v{c}ka c. 32, 10002 Zagreb, Croatia}
\date{}
\maketitle


\begin{abstract}
\noindent
We consider a scenario in which TeV-scale particles belonging to  weak-isospin
multiplets higher than triplets lead to novel seesaw mechanisms different from
conventional type I, II and III seesaw models.  Besides an appealing
testability of these mechanisms at the LHC, the model with Majorana quintuplets
with imposed discrete symmetry may provide viable dark matter candidate.
\end{abstract}

\section{Landmarks}

The lightness of the neutrinos, the lightness of the Higgs boson and the evidence for a Dark Matter (DM) have been the landmarks for 
extensions of the particle content of the Standard Model (SM).
Neutrino masses are the first tangible deviation from the SM so that the best motivated new particles appear in attempts 
to explain small neutrino masses.
The most popular explanation is the so-called seesaw mechanism in which extremely small masses of neutrinos arise on account of 
inverse proportionality to large masses of new, yet to be discovered particles.

In the simplest, Type I seesaw model~\cite{Minkowski:1977sc-etc}, new heavy particles are singlets under the Standard Model (SM) gauge group.
Devoid of SM charges they do not feel SM forces and resemble the elusive substance contemplated by Rudjer Bo\v{s}kovi\'c long ago.
In view of the 250th anniversary of the Venice-edition of his {\em Theoria}~\cite{Bos_en,Bos_hr}, it seems timely to recall
the prophetic thoughts in point No. 518  therein:\\
 {\em ``It could be possible to imagine that there is no force between some of existing species.
In this case, the substance of one kind might freely pass through that of another,
without any collision. In the same way, two of the species might have a common
force law with third species, without any force between the first two''.}

The Type I seesaw mediators, as a substance without SM charges, would ``pass through our matter without any collision''. 
In order to discover them,  it would be mandatory to assign to them a new charge (new gauge force) such as right-handed weak interaction.

Less elusive seesaw mediators seem to be those with nontrivial electroweak charges, as introduced in
the remaining two~\cite{Ma:1998dn} tree-level canonical seesaw mechanisms dubbed the type II~\cite{KoK-etc} and type III~\cite{Foot:1988aq}.
Their  scalar and fermion triplets allow for usual gauge invariant interactions with a neighbour  SM-doublet state.
This leads again to the Weinberg's~\cite{Weinberg:1979sa}  effective dimension-five operator  $LL\Phi\Phi$, and relegates the masses of exotic
seesaw mediators to remote $M \sim 10^{14}$ GeV of the GUT scale~\cite{Georgi:1974sy}.

It is conceivable that Nature simply does not provide particles which are needed for type I, II and III mechanisms and that the first exotic particles
we might encounter belong to the multiplets higher than the proposed triplets.
Such higher multiplets would be related to dimension $d=5+2n$ operators  $(LL\Phi\Phi)(\Phi\Phi)^n$ studied in \cite{Bonnet:2009ej}. The corresponding
light neutrino mass is given by the  seesaw formula $m_{\nu} \sim v_H (v_H/M)^{d-4}$, where $v_H$= 246 GeV is the vev of the SM higgs.
Accordingly, dimension-nine operators naturally  correspond to TeV-scale new particles, within the discovery reach of the LHC.

\vspace{.5cm}

\section{Model with Dirac lepton quintuplet}

In order to add something new to canonical tree-level seesaw mechanisms one can employ vectorlike fermionic multiplets with non-zero hypercharge. 
To generate a tree-level seesaw diagram, the hypercharge non-zero Dirac leptons have to be in conjunction with additional scalar fields.

Let us first systematize possible realizations of such tree-level mechanism~\cite{Picek:2009is,Kumericki:2011hf} and then focus to dimension-nine model realized
with Dirac lepton quintuplet $\Sigma_{L,R} \sim (1, 5, 2)$.
Since the sought-after higher-dimension operator can be relevant only in the absence of possible dimension-five operator, we forbid 
the existence of the states which may generate conventional seesaw mechanisms: a scalar triplet $\Delta\sim(1,3,2)$ generating type II seesaw, and a fermion singlet $N_R\sim(1,1,0)$ or triplet $N_R\sim(1,3,0)$  generating type I and III seesaw, respectively.
Then, we are restricted to two options for seesaw mediators underlined in Table~\ref{table_conjunct}:  the triplet fermions introduced in~\cite{Babu:2009aq} and the quintuplet fermions proposed in~\cite{Picek:2009is,Kumericki:2011hf}, in our focus here.\\

\begin{table}[h]
\begin{center}
\begin{tabular}{ccccc}
\hline
Seesaw Type    & Exotic Fermion           & Exotic Scalar                       & Scalar Coupling        & $m_\nu$ at     \\ \hline
Type I         & $N_R\sim(1,0)$           & -                                      & -                     & dim 5          \\
Type II        & -                        & $\Delta\sim(3,2)$                        & $\mu \Delta H H$        & dim 5          \\
Type III       & $N_R\sim(3,0)$           & -                                      & -                     & dim 5          \\ \hline
Conjunct    & Exotic Fermion           & Exotic Scalars                       & Scalar - Higgs        & $m_\nu$ at     \\
    Mediator       &  Pair                    & $\Phi_1,\ \Phi_2$                   & Couplings        &      \\ \hline
doublet        & $\Sigma_{L,R}\ (2,1)$  & $(3,-2),\   (3,0)$  & $\mu_{1,2} \Phi_{1,2} H H$ & dim 5          \\
\textbf{triplet}        & $\Sigma_{L,R}\ (3,2)$  & $(4,-3),\   (2,-1)$   & $\lambda_1\Phi_1 H H H$       & \textbf{dim 7}          \\
quadruplet    & $\Sigma_{L,R}\ (4,1)$  & $(3,-2),\   (3,0)$  & $\mu_{1,2} \Phi_{1,2} H H$ & dim 5          \\
\textbf{quintuplet}    & $\Sigma_{L,R}\ (5,2)$  & $(4,-3),\   (4,-1)$   & $\lambda_{1,2}\Phi_{1,2} H H H$   & \textbf{dim 9}          \\\hline
\end{tabular}
\caption{The assignments of electroweak charges for exotic particles in case of Dirac seesaw mediators leading to the tree-level operators up to dimension nine.}
\label{table_conjunct}
\end{center}
\end{table}
The model with Dirac lepton quintuplet starts from three generations of SM leptons $L_L$ and $l_R$ completed with $n_\Sigma$ isospin $T=2$ vectorlike quintuplets 
with hypercharge two, $\Sigma_{L,R} \sim (1,5,2)$. Also, besides the SM Higgs doublet $H$ there are two additional scalar quadruplets $\Phi_1$ and $\Phi_2$ 
transforming as $(1,4,-3)$ and $(1,4,-1)$,
\begin{equation}\label{quintuplet_dir}
\Sigma_{L,R}=\left(\begin{array}{l}
\Sigma^{+++}\\\Sigma^{++}\\\Sigma^{+}\\\Sigma^{0}\\\Sigma^{-}
\end{array}\right)_{L,R}; \ \ \ \
\Phi_1=\left(\begin{array}{l}
\phi_1^0\\\phi_1^-\\\phi_1^{--}\\\phi_1^{---}
\end{array}\right)\ \ , \ \ \ \
\Phi_2=\left(\begin{array}{l}
\phi_2^+\\\phi_2^0\\\phi_2^{-}\\\phi_2 ^{--}
\end{array}\right)\ .
\end{equation}
Gauge invariant Lagrangian includes the Yukawa and Dirac mass terms 
\begin{eqnarray}\label{lagrangian_dirac}
\mathcal{L}&=& \overline{\Sigma_L}i D_\mu \gamma^\mu\Sigma_L + \overline{\Sigma_R}iD_\mu \gamma^\mu\Sigma_R
-\overline{\Sigma_R}M_\Sigma \Sigma_L -\overline{\Sigma_L}M_\Sigma^\dagger \Sigma_R \nonumber \\
&+& \left( \overline{\Sigma_R}Y_1L_L\Phi_1^* + \overline{(\Sigma_L)^c} Y_2 L_L \Phi_2+ \mathrm{H.c.} \right)\ .
\end{eqnarray}
The scalar potential contains renormalizable terms relevant for our mechanism
\begin{eqnarray}\label{potential_dirac}
V(H, \Phi_1, \Phi_2) &\sim& -\mu_H^2 H^\dagger H + \mu^2_{\Phi_1} \Phi^\dagger_1 \Phi_1+ \mu^2_{\Phi_2} \Phi^\dagger_2 \Phi_2 + \lambda_H (H^\dagger H )^2 \nonumber \\
 &+& \{ \lambda_1 \Phi^*_1 H^* H^* H^* + \mathrm{H.c.} \} + \{ \lambda_2 \Phi^*_2 H H^* H^* + \mathrm{H.c.} \} \nonumber\\
 &+& \{ \lambda_3 \Phi^*_1 \Phi_2 H^* H^* + \mathrm{H.c.} \} \ ,
\end{eqnarray}
where $\lambda_1$ and $\lambda_2$ terms induce vevs for the scalar quadruplets, which together with the Yukawa couplings 
define the matrix-valued couplings  $V_1$ and $V_2$
\begin{equation}\label{vev_dir}
    v_{\Phi_1} \simeq -\lambda_1 \frac{v_H^3}{\mu^2_{\Phi_1}} \ \  ,\ \ 
v_{\Phi_2} \simeq -\lambda_2 \frac{v_H^3}{\mu^2_{\Phi_2}} \ \ ; \ \
V_1 \sim  Y^\dagger_1 \frac{v_{\Phi_1}} {M_\Sigma} \ \ , \ \ V_2 \sim  Y^\dagger_2 \frac{v^*_{\Phi_2}} {M_\Sigma} \ \ .
\end{equation}

\subsection{Neutrino masses}

\begin{figure}[h]
\begin{center}
\includegraphics[scale=1.1]{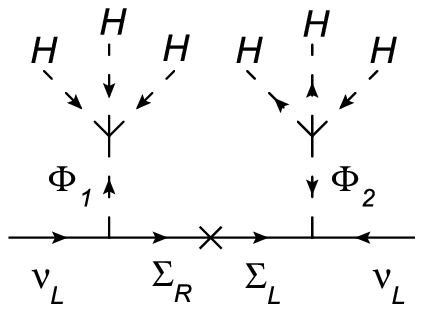} \hspace{0.9cm}
\includegraphics[scale=1.1]{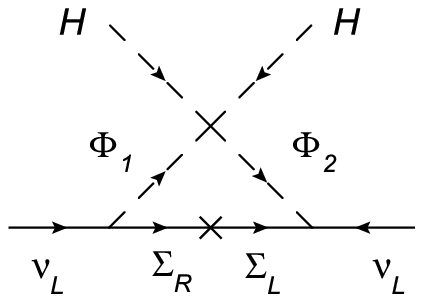}
\caption{Tree-level diagram contribution (LHS) and one-loop diagram contribution (RHS) in Eq.~(\ref{tree_dir}).}
\label{dim9_dir}
\end{center}
\end{figure}

The induced vevs $v_{\Phi_1}$ and $v_{\Phi_2}$ and the Yukawa terms in Eq.~(\ref{lagrangian_dirac}) lead to the mass terms connecting the SM lepton doublet with new Dirac quintuplet lepton. Three neutral left-handed fields $\nu_L$, $\Sigma_L^{0}$ and $(\Sigma_R^{0})^c$ span the symmetric neutral mass matrix as follows:
\begin{eqnarray}
\mathcal{L}_{\nu \Sigma^0} =  \, -\frac{1}{2}
\left(  \overline{(\nu_L)^c}  \; \overline{(\Sigma_L^0)^c} \; \overline{\Sigma_R^0} \right)
\left( \! \begin{array}{ccc}
0 & m_2^T & m_1^T \\
m_2 & 0 & M_{\Sigma}^T \\
m_1 & M_{\Sigma} & 0
\end{array} \! \right) \,
\left( \!\! \begin{array}{c} \nu_L \\ \Sigma_L^0 \\ (\Sigma_R^0)^c \end{array} \!\! \right)
\; + \mathrm{H.c.}\ .
\label{matrix_neutral_dir}
\end{eqnarray}
The diagonalization of this mass matrix leads to tree-level contribution to the light neutrino mass, and the quartic coupling $\lambda_3$ in Eq.~(\ref{potential_dirac}) gives the loop contribution to the neutrino masses
\begin{equation}\label{tree_dir}
    m_{\nu}^{tree} \sim \frac {Y_1 Y_2\ \lambda_1\lambda_2\ v_H^6} {M_{\Sigma}\ \mu^2_{\Phi_1}\ \mu^2_{\Phi_2}} \ \ \ \ \ \ , \ \ \ \ \ \ 
    m_{\nu}^{loop} \sim \frac {Y_1 Y_2\ \lambda_3\ v_H^2} {16 \pi^2 \ M_{\Sigma}}
\end{equation}
corresponding to dimension-nine tree-level seesaw mechanism and to dimension-five radiative mechanism displayed on LHS and RHS of Fig.~\ref{dim9_dir}, respectively.

\subsection{Production and decays of Dirac quintuplet leptons at the LHC}

The production channels of the heavy quintuplet leptons in proton-proton collisions, dominated by the quark-antiquark annihilation
\begin{displaymath}
q + \bar{q} \to A \to \Sigma + \bar{\Sigma}\;, \qquad A = \gamma, Z, W^\pm \;,
\end{displaymath}
are determined entirely by gauge couplings of neutral and charged gauge bosons.
\begin{figure}[h]
\begin{center}
\includegraphics[scale=0.64]{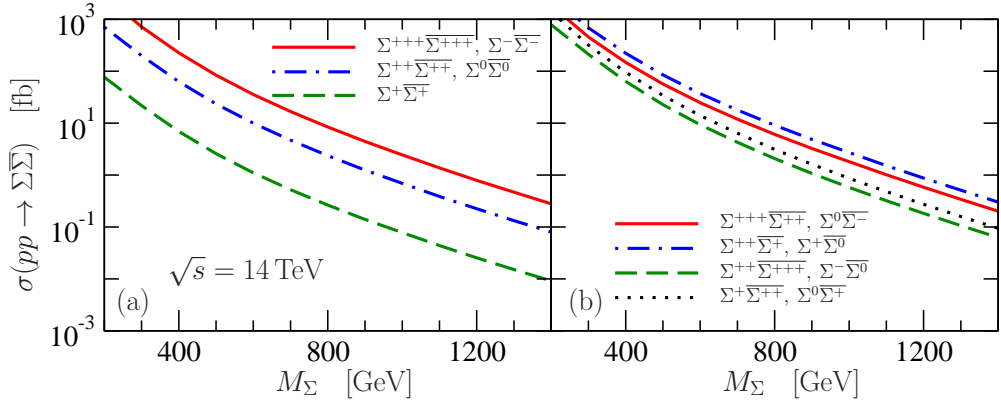}
\caption{The cross sections for production of Dirac quintuplet lepton pairs on LHC
proton-proton collisions at designed $\sqrt{s}=14\, {\rm TeV}$ via
neutral $\gamma, Z$ (a) and charged $W^{\pm}$ currents (b),
in dependence on the heavy quintuplet mass $M_\Sigma$.}
\label{production14_dir}
\end{center}
\end{figure}
The cross sections for proton-proton collisions are presented for designed $\sqrt{s}=14\,{\rm TeV}$ on Fig.~\ref{production14_dir}. Thereby we distinguish 
the production via neutral currents shown on LHS and via charged currents shown on RHS of Fig.~\ref{production14_dir}.  By testing the heavy lepton production 
cross sections one can hope to identify the quantum numbers of Dirac quintuplet particles, but in order to confirm their relation to neutrinos one has to study their decays.
Provided that the exotic scalar states are slightly heavier than the exotic leptons, the exotic scalars will not appear in the final states in  heavy lepton decays. 
\\
\begin{table}[h]
\centering
\begin{tabular}{|c||c|c|c|c|} \hline
                         &  $\overline{\Sigma^+}\to \ell^-Z^0$   &   $\overline{\Sigma^0}\to \ell^+W^-$   &   $\overline{\Sigma^0}\to \ell^-W^+$   &  $\overline{\Sigma^-}\to \ell^+Z^0$ \\\hline
\hline
$\Sigma^+\to \ell^+Z^0$   &  $\ell^+ \ell^-Z^0 Z^0 $   &   $\ell^+ \ell^+Z^0 W^- $     &   $\ell^+ \ell^-Z^0 W^+ $     &  -  \\\hline
$\Sigma^0\to \ell^-W^+$   &  $\ell^- \ell^-W^+ Z^0 $   &   $\ell^- \ell^+W^+ W^- $     &   $\ell^- \ell^-W^+ W^+ $     &  $\ell^- \ell^+W^+ Z^0 $ \\\hline
$\Sigma^0\to \ell^+W^-$   &  $\ell^+ \ell^-W^- Z^0 $   &   $\ell^+ \ell^+W^- W^- $     &   $\ell^+ \ell^-W^- W^+ $     &  $\ell^+ \ell^+W^- Z^0 $ \\\hline
$\Sigma^-\to \ell^-Z^0$   &  -    &   $\ell^- \ell^+Z^0 W^- $     &   $\ell^- \ell^-Z^0 W^+ $     &  $\ell^- \ell^+Z^0 Z^0 $ \\\hline
\end{tabular}
\caption{\footnotesize Decays to SM particles including same sign dilepton events}
\label{table_decay_dir}
\end{table}
\\

\textbf{Pointlike decays}
$\ $to gauge bosons and SM leptons are experienced by four lowest states out of the five $\Sigma$-states properly ordered in Eq.~(\ref{quintuplet_dir}). In Table~\ref{table_decay_dir} we list all possible events coming from the decays of the neutral and singly-charged Dirac quintuplet states to the SM charged leptons. This includes the same-sign dilepton events as a distinguished signature at the LHC.

The partial decay width of $\Sigma^{++}$ state decaying exclusively via a charged current is 
\begin{equation}\label{width++_dir}
    \Gamma(\Sigma^{++}\to \ell^+W^+)= {g^2\over 32\pi} \Big| \sqrt3 V_2^{\ell \Sigma}\Big|^2 {M_\Sigma^3\over M_W^2}\left(1-{M_W^2\over
M_\Sigma^2}\right)^2\left(1+2{M_W^2\over M_\Sigma^2}\right).
\end{equation}
There is no such pointlike decay of the triply-charged $\Sigma^{+++}$ state which has other interesting decays presented below.
\\

\textbf{Cascade decays}
$\ $ $\Sigma^i \to \Sigma^j \pi^+$ and $\Sigma^i \to \Sigma^j l^+\nu$ are suppressed by small mass differences, except for $\Sigma^{+++}$ decays. 
These decays will serve as the referent decays for the golden decay mode of the triply-charged state.
\\

\begin{figure}[h]
\begin{center}
\includegraphics[scale=0.52]{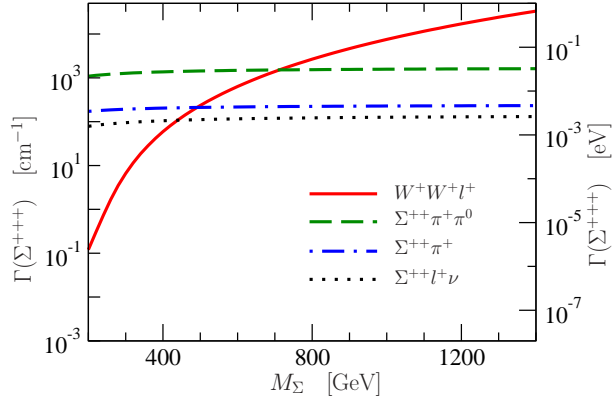}
\caption{Selected partial decay widths of $\Sigma^{+++}$ Dirac quintuplet lepton for $|V^{l\Sigma}_2| = 10^{-6} \sqrt{{800 \over M_{\Sigma} \rm{(GeV)}}}$ in dependence of heavy quintuplet mass $M_{\Sigma}$.}
\label{decay+++_dir_800}
\end{center}
\end{figure}

\textbf{Golden decay}
$\ $mode $\Sigma^{+++} \to W^+ W^+ l^+$ given by partial width plotted on Fig.~\ref{decay+++_dir_800}, $\Gamma(\Sigma^{+++} \to W^+ W^+ l^+) \sim M_\Sigma^5 / M_W^4$ in the limit $M_\Sigma \gg M_W$, is governed by the same mixing factor $V_2$ which determines the  $\Sigma^{++}$ decay in Eq.~(\ref{width++_dir}). On the same figure we plot the partial widths for other decays of $\Sigma^{+++}$.
For $5\ \rm{fb^{-1}}$ of integrated luminosity of 2011 LHC run at $\sqrt{s}=7$ TeV and $21\ \rm{fb^{-1}}$ of integrated luminosity of 2012 LHC run at $\sqrt{s}=8$ TeV, there could be 8000 $\Sigma$-$\overline \Sigma$ pairs in total produced for $M_\Sigma=400 \,\rm{GeV}$. There are 3300 triply-charged $\Sigma^{+++}$ or $\overline{\Sigma^{+++}}$ fermions among them, resulting in $\sim 300$ golden decays $\Sigma^{+++}(\overline{\Sigma^{+++}}) \to W^{\pm} W^{\pm} l^{\pm}$. This makes the model with Dirac quintuplet falsifiable at the LHC. A mere nonobservance of triply-charged Dirac fermions would put a study of Majorana quintuplet in a forefront.

\vspace{.5cm}

\section{Model with Majorana Quintuplets}

This model~\cite{Kumericki:2012bh} adds to SM fermions three generations of hypercharge zero lepton quintuplets $\Sigma_R=(\Sigma_R^{++},\Sigma_R^{+},\Sigma_R^{0},\Sigma_R^{-},\Sigma_R^{--})$, transforming as $(1,5,0)$ under the SM gauge group. Also, in addition to SM Higgs doublet $H= (H^+, H^0)$ there is a scalar quadruplet $\Phi=(\Phi^{+},\Phi^{0},\Phi^{-},\Phi^{--})$ transforming as $(1,4,-1)$.\\
The gauge invariant and renormalizable Lagrangian involving these new fields reads
\begin{equation}\label{lagrangian_majorana}
   \mathcal{L} = \overline{\Sigma_R} i \gamma^\mu D_\mu \Sigma_R + (D^\mu \Phi)^\dag (D_\mu \Phi) -
   \big(\overline{L_L} Y \Phi  \Sigma_R + {1 \over 2} \overline{(\Sigma_R)^C} M \Sigma_R + \mathrm{H.c.}\big)
- V(H,\Phi)  \ .
\end{equation}
Here, $Y$ is the Yukawa-coupling matrix and $M$ is the mass matrix of the heavy leptons, which contains the terms containing two charged Dirac fermions and one neutral Majorana fermion
\begin{equation}\label{fermions_maj}
   \Sigma^{++} = \Sigma^{++}_R + \Sigma^{--C}_R\ ,\ \Sigma^+ = \Sigma^+_R - \Sigma^{-C}_R\ ,\ \Sigma^0
= \Sigma^0_R + \Sigma^{0C}_R\ .
\end{equation}

The scalar potential, assuming real quartic couplings, has the gauge invariant form
\begin{eqnarray}\label{scalarpot}
\nonumber  V(H,\Phi) &=& -\mu_H^2 H^\dagger H + \mu_\Phi^2 \Phi^\dagger \Phi + \lambda_1 \big( H^\dagger H \big)^2 + \lambda_2 H^\dagger H \Phi^\dagger \Phi + \lambda_3 H^* H \Phi^* \Phi \\
\nonumber   &+& \big( \lambda_4 H^* H H \Phi+ \mathrm{H.c.} \big) + \big( \lambda_5 H H \Phi \Phi + \mathrm{H.c.} \big) + \big( \lambda_6 H \Phi^* \Phi \Phi+ \mathrm{H.c.} \big)\\
            &+& \lambda_7 \big( \Phi^\dagger \Phi \big)^2 + \lambda_8 \Phi^* \Phi \Phi^* \Phi  \ .
\end{eqnarray}
The presence of the $\lambda_4$ term leads to the induced vev $v_\Phi \sim - \lambda_4 v_H^3 / \mu_\Phi^2$.

\subsection{Neutrino masses}

\begin{figure}[h]
\begin{center}
\includegraphics[scale=1.1]{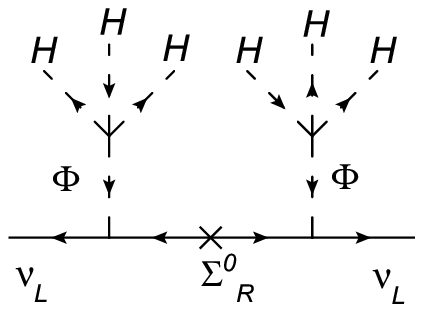} \hspace{0.9cm}
\includegraphics[scale=1.1]{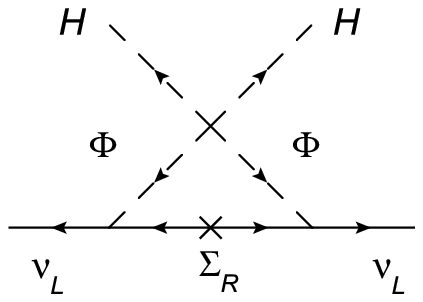}
\caption{\small Tree-level diagram corresponding to dimension-nine operator and one-loop diagrams corresponding to dimension-five operator.}
\label{dim9op}
\end{center}
\end{figure}

The vev $v_\Phi$ generates a Dirac mass term connecting $\nu_L$ and $\Sigma_R^0$, a nondiagonal entry in
the mass matrix for neutral leptons given by
\begin{eqnarray}
\mathcal{L}_{\nu \Sigma^0} =  \, -\frac{1}{2}
\left(  \overline{\nu_L}  \; \overline{(\Sigma_R^0)^C} \right)
\left( \! \begin{array}{cc}
0 & \frac{1}{\sqrt{2}} Y v_\Phi \\
\frac{1}{\sqrt{2}} Y^T v_\Phi & M
\end{array} \! \right) \,
\left( \!\! \begin{array}{c} (\nu_L)^c \\ \Sigma_R^0 \end{array} \!\! \right)
\; + \mathrm{H.c.}\ .
\label{neutral_mass_matrix}
\end{eqnarray}
After diagonalizing this mass matrix the light neutrinos acquire the Majorana mass contribution corresponding to tree-level diagram displayed on LHS of Fig.~\ref{dim9op}. Simultaneously, the quartic $\lambda_5$ term in Eq.~(\ref{scalarpot}) generates the one-loop diagram displayed on RHS of Fig.~\ref{dim9op}.
These two contributions added together give the light neutrino mass matrix
\begin{eqnarray}\label{mnu}
  (m_\nu)_{ij} &=& (m_\nu)_{ij}^{tree}+(m_\nu)_{ij}^{loop} \nonumber\\
               &=& \frac{-1}{6} (\lambda^*_4)^2 \frac{v^6}{\mu_\Phi^4} \sum_k {Y_{ik} Y_{jk} \over M_k}
+ {-5 \lambda_5^* v^2 \over 24 \pi^{2}}
\sum_k {Y_{ik} Y_{jk} M_k \over m_\Phi^{2} - M_k^{2}} \left[
1 - {M_k^{2} \over m_\Phi^{2}-M_k^{2}} \ln {m_\Phi^{2} \over M_k^{2}}  \right] \,. \nonumber\\
\end{eqnarray}

\subsection{Production and decays of Majorana quintuplet leptons at the LHC}

The production channels of heavy quintuplet leptons in proton-proton collisions are
dominated by the quark-antiquark annihilation via neutral and charged gauge bosons.

\begin{figure}
\begin{center}
\includegraphics[scale=0.8]{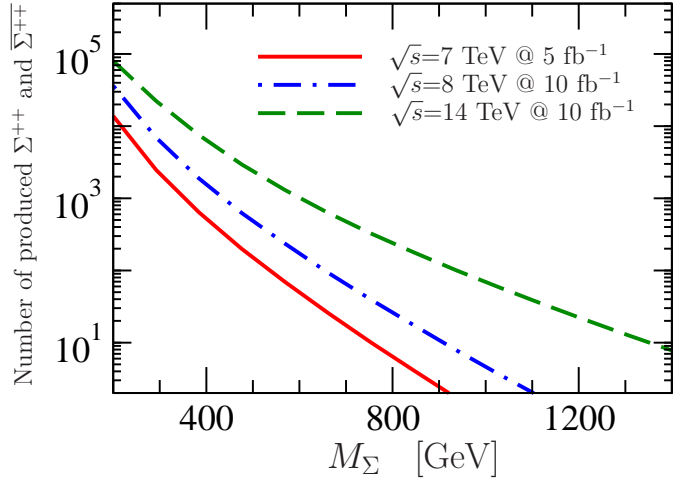}
\caption{Number of $\Sigma^{++}$ and
$\overline{\Sigma^{++}}$ particles produced for three
characteristic LHC collider setups, in dependence on
the heavy lepton mass $M_{\Sigma}$.}
\label{fig:LHCproduction}
\end{center}
\end{figure}
On Fig. \ref{fig:LHCproduction} we plot the expected number of produced $\Sigma^{++}$ and $\overline{\Sigma^{++}}$ particles for three
characteristic collider setups. In particular, for $M_\Sigma = 400 \, \rm{GeV}$ and $5\ \rm{fb^{-1}}$ of integrated luminosity of 2011 LHC run at $\sqrt{s}=7$ TeV, and $21\ \rm{fb^{-1}}$ of integrated luminosity of 2012 LHC run at $\sqrt{s}=8$ TeV, there should be about 3200 doubly-charged $\Sigma^{++}$ or $\overline{\Sigma^{++}}$ fermions produced. In total, there should be 4000 $\Sigma - \overline{\Sigma}$ pairs produced.

In order to confirm the relation of the quintuplet particles to neutrinos one has to study their decays. We list the representative final states of these decays in Table~\ref{events}, which includes same-sign dilepton events as distinguished signatures at the LHC.
\begin{table}[h]
\centering
\begin{tabular}{|c||c|c|c|c|} \hline
 &  $\overline{\Sigma^{++}}\to \ell^-W^- $   &
 $\overline{\Sigma^+}\to \ell^-Z^0 $   &   $\Sigma^0\to \ell^+W^- $   &
 $\Sigma^0\to \ell^-W^+ $ \\
 &  $(0.66)$   &
 $(0.06)$   &   $(0.30)$   &
 $(0.30)$ \\\hline
\hline
$\Sigma^{++}\to \ell^+W^+ $   &  $\ell^+ \ell^-W^+ W^- $   &
 $\ell^+ \ell^- W^+ Z^0 $     &   -     &  -  \\
$ (0.66)$   &  $ (0.44)$   &
 $ (0.04)$     &   -     &  -  \\\hline
$\Sigma^+\to \ell^+Z^0 $   &  $\ell^+ \ell^- Z^0 W^-  $   &
   $\ell^+ \ell^- Z^0 Z^0  $     &   $\ell^+ \ell^+ Z^0 W^-  $     &
  $\ell^+ \ell^- Z^0 W^+  $ \\
$ (0.06)$   &  $(0.04) $   &
   $ (0.004) $     &   $ (0.02) $     &
  $ (0.02) $ \\\hline
$\Sigma^0\to \ell^-W^+ $   &  -    &   $\ell^- \ell^-W^+ Z^0  $     &
-     &  - \\
$(0.30)$   &  -    &   $(0.02) $     &   -     &  - \\\hline
$\Sigma^0\to \ell^+W^- $   &  -    &   $\ell^+ \ell^-W^- Z^0  $     &
-     &  - \\
$(0.30)$   &  -    &   $(0.02) $     &   -     &  - \\\hline
\end{tabular}
\caption{\footnotesize Decays of exotic leptons to SM charged leptons,
including multi-lepton and same-sign dilepton events, together with
their branching ratios (restricted to $l=e, \mu$ and for $M_\Sigma =
400\,{\rm GeV}$).}
\label{events}
\end{table}
The distinctive signatures could come from doubly-charged components of the fermionic quintuplets.
The signals which are good for the discovery correspond to relatively high signal rate and small SM background. Two promising classes of events contain $\Sigma^+$ decaying to $e^+$ or $\mu^+$ lepton and $Z^0 \to (\ell^+ \ell^-, q \bar{q})$ resonance  helping in $\Sigma^+$ identification:
\begin{displaymath}
(i) \qquad p \, p \to \Sigma^+  \, \overline{\Sigma^0} \to (\ell^+Z^0) \, (\ell^+W^-) \;,
\end{displaymath}
the LNV event having 0.7 fb with the same-sign dilepton state, which is nonexistent in the SM and thus devoid of the SM background;
\begin{displaymath}
(ii) \qquad p \, p \to \Sigma^{++}  \, \overline{\Sigma^+} \to (\ell^+W^+) \, (\ell^-Z^0) \;,
\end{displaymath}
having relatively high signal rate of 1.1 fb with respect to the SM background of 0.8 fb.

\subsection{Higgs diphoton-decay induced by doubly-charged scalars}

The recent discovery of a Higgs-like boson at the Large Hadron Collider (LHC)
may mark an onset of a new set of particles with mass at the  100 GeV scale. Simultaneously, a ``virtual physics'' in the LHC era may
have important effects in rare, loop-induced processes.\\
We have considered~\cite{Picek:2012ei} the possible enhancement of the $h \rightarrow \gamma \gamma$ decay rate through extra contributions with charged components of scalar quadruplet $\Phi$ running in the loops, additional to dominant SM contributions from the $W$ boson and top quark loops. The analytic expression for the diphoton $h \rightarrow \gamma \gamma$ partial width reads~\cite{Carena:2012xa}
\begin{equation}
\label{W-t-S-loop}
\Gamma(h\to \gamma \gamma)=\frac{\alpha^2 m_h^3}{256 \pi^3 v_H^2}\left|A_1(\tau_W)+ N_c Q_t^2  A_{1/2}(\tau_t)
+  N_{c,S} Q_S^2 \frac{c_S}{2} \frac{v_H^2}{m_S^2} A_0(\tau_S)\right |^2 \ ,
\end{equation}
where the three contributions corresponding to $\tau_i\equiv 4m_i^2/m_h^2$ ($i=W, t, S$) refer to  spin-1 ($W$ boson), spin-1/2 (top quark) and  charged spin-0 particles in the loop.
Following~\cite{Carena:2012xa} we define the enhancement factor with respect to the SM decay width
\begin{equation}
    R_{\gamma\gamma} = \left| 1+  \sum_{S=\Phi^{+},\Phi^{-},\Phi^{--}} Q_S^2 \frac{c_S}{2} \frac{v_H^2}{m_{S}^2}\frac{A_0(\tau_{S})}{ A_1(\tau_W)+ N_c Q_t^2 \, A_{1/2}(\tau_t)}\right|^2 \ .
\end{equation}
This enhancement is dominated by the lightest charged scalar $\Phi^{--}$ and the value $R_{\gamma\gamma}=2$ can be achieved up to $m(\Phi^{--})=280$ GeV, and the value $R_{\gamma\gamma}=1.25$ up to $m(\Phi^{--})=520$ GeV.

Another loop mediated Higgs decay sensitive to new charged particles is $h \rightarrow Z \gamma$, where we obtain a moderate suppression of the $h \rightarrow Z \gamma$ decay rate in the region of the parameter space where the $h \rightarrow \gamma \gamma$ decay rate is enhanced.

\vspace{.5cm}

\section{Conclusions}

The need for neutrino masses modifies the ``old'' SM to a ``new $\nu$SM'' where, in order to understand the lightness of active neutrinos, one commonly introduces
new heavy degrees of freedom. While no evidence for such new particles has been found so far at the LHC, evidences from astrophysics and cosmology point at 23$\%$ of the energy density of the Universe provided by DM~\cite{Nakamura:2010zzi}. Notably, appealing DM candidates are heavy neutrino-like weakly interacting massive particles (WIMPs).

Bearing in mind an accidental stability of our ordinary matter, it is tempting to imagine a setup in which DM is also accidentally stable.
The so-called minimal dark matter model (MDM)~\cite{Cirelli:2005uq} is probably the simplest such scenario. In this setup the stability can be guaranteed 
for a neutral component of a large enough fermion or scalar $SU(2)_L$ multiplet,
which can not form $SU(2)_L$ invariant renormalizable (or lowest nonrenormalizable, dimension-five) interaction-terms with the SM multiplets. For a fermion candidate it would be
the quadruplet (quintuplet) and higher, and for a scalar candidate the sextuplet  and higher. Since the neutral (DM) component cannot have the tree level interactions to the Z boson, both the hypercharge $Y$ and the third isospin-component $T_3$ must vanish. This is possible only for odd multiplets: fermion quintuplet or higher, and scalar septuplet or higher.

 When employing the fermion quintuplets as seesaw mediators, like in two models presented here, they are in conjunction with appropriate scalar quadruplets. 
This destroys the stability which the MDM Majorana quintuplet may possess in isolation.

By exploring the conditions under which the quintuplets $\Sigma_R \sim (1, 5, 0)$ could simultaneously generate neutrino masses and provide a stable DM candidate,
the paper~\cite{Cai:2011qr} raised the hope that the Majorana quintuplet in conjunction with a scalar septuplet can do the job, leading to radiative neutrino mass 
mechanism (R$\nu$MDM) with an automatic $Z_2$ symmetry. However, we have demonstrated~\cite{Kumericki:2012bf}  that this is still not possible without 
imposing the discrete  $Z_2$ symmetry by hand. This brings us back  to the model~\cite{Kumericki:2012bh} with a Majorana quintuplet and a scalar quadruplet as a more minimal option.
The mass of $\Sigma^0$, as the DM particle protected by the $Z_2$ symmetry, is fixed by the relic abundance to the value~\cite{Cirelli:2005uq} $M_\Sigma \approx 10\ \rm{TeV}$. The choice $\lambda_5=10^{-7}$ gives enough suppression to lead to small neutrino masses with large Yukawas, $Y \sim 0.1$. In this part of the parameter space the model could have interesting LFV effects like in~\cite{Cai:2011qr}. 

Finally, we should keep in mind the possibility that DM may belong to a setup not very different from fine-tuned (anomaly-free) set of ordinary SM particles. 
A mere mismatch in the charges of the SM and DM particles explains a separate stability of two parallel worlds, 
and the adventure of explaining  neutrino masses may for the first time reveal the charges of some of the DM species.

\section*{Acknowledgments}
This work is supported by the Croatian Ministry  of Science, Education and
Sports under Contract No. 119-0982930-1016.

\end{document}